\documentclass[twocolumn,reprint,superscriptaddress,aps,pra]{revtex4-2}
\usepackage{amsmath}
\usepackage{amssymb}
\usepackage{units}
\usepackage{graphicx,subfigure}
\usepackage{mathrsfs}
\usepackage{bm}
\usepackage{textcomp}
\usepackage{url}
\usepackage{dsfont} 
\usepackage{braket}
\usepackage[compact]{titlesec}
\usepackage[colorlinks=true,linkcolor=magenta,citecolor=blue]{hyperref}
\usepackage{todonotes}
\usepackage{empheq}
\usepackage{multirow}
\usepackage{ulem}
\usepackage{color}
\begin{document}
\title{Generating Grating in Cavity Magnomechanics}
\author{Wenzhang Liu}
\email{oskarliu@stu.xjtu.edu.cn}
\affiliation{Ministry of Education Key Laboratory for Nonequilibrium Synthesis and Modulation of Condensed Matter,
Shaanxi Province Key Laboratory of Quantum Information and Quantum Optoelectronic Devices, School of Physics, Xi’an Jiaotong University, Xi’an 710049, China}
\author{Muqaddar Abbas}
\email{muqaddarabbas@xjtu.edu.cn}
\affiliation{Ministry of Education Key Laboratory for Nonequilibrium Synthesis and Modulation of Condensed Matter,
Shaanxi Province Key Laboratory of Quantum Information and Quantum Optoelectronic Devices, School of Physics, Xi’an Jiaotong University, Xi’an 710049, China}
\author{Seyyed Hossein Asadpour}
\affiliation{School of Physics, Institute for Research in Fundamental Sciences (IPM), Tehran 19395-5531, Iran}
\author{Hamid R. Hamedi}
\email{hamid.hamedi@tfai.vu.lt}
\affiliation{Institute of Theoretical Physics and Astronomy, Vilnius University, Sauletekio 3, Vilnius 10257, Lithuania}
\author{Pei Zhang}
\email{zhangpei@mail.ustc.edu.cn}
\affiliation{Ministry of Education Key Laboratory for Nonequilibrium Synthesis and Modulation of Condensed Matter,
Shaanxi Province Key Laboratory of Quantum Information and Quantum Optoelectronic Devices, School of Physics, Xi’an Jiaotong University, Xi’an 710049, China}
\author{Barry C. Sanders}
\email{sandersb@ucalgary.ca}
\affiliation{Institute for Quantum Science and Technology, University of Calgary, Alberta T2N~1N4, Canada}
\date{\today}
\begin{abstract}
We investigate the phenomenon of magnomechanically induced grating (MMIG) within a cavity magnomechanical system, comprising magnons (spins in a ferromagnet, such as yttrium iron garnet), cavity microwave photons, and phonons [\textit{J. Li, S.-Y. Zhu, and G. S. Agarwal, Phys. Rev. Lett. \textbf{121}, 203601 (2018)}]. By applying an external standing wave control, we observe modifications in the transmission profile of a probe light beam, signifying the presence of MMIG. Through numerical analysis, we explore the diffraction intensities of the probe field, examining the impact of interactions between cavity magnons, magnon-phonon interactions, standing wave field strength, and interaction length. MMIG systems leverage the unique properties of magnons, and collective spin excitations with attributes like long coherence times and spin-wave propagation. These distinctive features can be harnessed in MMIG systems for innovative applications in information storage, retrieval, and quantum memories, offering various orders of diffraction grating.
\end{abstract}
\maketitle
\section{Introduction}
\label{Sec:intro}
Electromagnetically induced transparency (EIT) is a well-established phenomenon wherein a typically opaque medium becomes transparent when exposed to a specific type of electromagnetic radiation~\cite{fleischhauer2005electromagnetically}. This nonlinear optical effect proves valuable in enhancing interactions while minimizing destructive processes, particularly photon absorption. Consequently, systems employing EIT exhibit the potential for facilitating long-distance quantum communication~\cite{chang2014quantum}.

When the control beam is configured as a standing-wave field, EIT can be harnessed to create a diffraction grating, giving rise to another phenomenon known as electromagnetically induced grating (EIG)~\cite{ling1998electromagnetically}. This configuration allows for the creation of both spatially absorbing (amplitude) and dispersion (phase) gratings in the sample, offering greater flexibility than classical optical gratings. EIG has a wide range of uses. For example, the arrangement of photonic gap bands may be changed through a grating formed by an optically generated lattice~\cite{kuang2010tunable, bajcsy2003stationary}.

EIG can be employed to generate an electromagnetically induced Talbot effect~\cite{wen2017two}, which proves highly advantageous for imaging mutually exclusive ultra-cold atoms. The tunability of gratings for diffraction in EIG opens up promising applications in various fields~\cite{de2010electromagnetically, zhai2001optical}. Subsequent experimental verifications of EIG were conducted in both cold~\cite{zhang2011four, mitsunaga1999observation} and hot~\cite{cardoso2002electromagnetically} atomic samples.

In addition, a lot of research has recently been done on mechanical oscillators as transducers that mediate the conversion of coherent signals across various systems~\cite{RevModPhys.86.1391}. Radiation force~\cite{PhysRevLett.95.033901, PhysRevLett.97.243905}, electrostatic force~\cite{articleee}, as well as piezoelectric element force~\cite{articlebb} has all been employed to couple phonons with optically or microwave photons. Such interaction processes result in the rapid emergence of a wide range of cavity electro- and optomechanical systems~\cite {PhysRev.124.1866}, although they all lack adequate tunability.

The magnetostrictive force~\cite{Butcher_2023} offers an additional method for coupling a different information carrier magnon with a phonon. Magnon is a collective excitation of magnetization, and its frequency may be changed at any time by altering the biased magnetic field~\cite{PhysRevLett.1.241}. Because magnetostrictive contact is weak in most dielectric or metallic materials, that is easy to ignore it while processing data. Due to the dominance of the magnetostrictive force in magnetic materials, a very flexible hybrid system for coherent information processing may be developed~\cite{PhysRevX.13.021016}.

A microwave cavity combined with a ferromagnetic material, such as a yttrium iron garnet (YIG) sphere, is the most common physical implementation of a cavity magnonic system. In recent decades, this system has attracted a lot of attention and shown extraordinary performance. YIG has garnered significant attention due to its distinctive characteristics, including a high spin density and an exceptionally low loss rate~\cite{zhang2016cavity, lachance2019hybrid}. Previously, YIG has found applications in magnetic storage~\cite{serga2010yig}, spintronics~\cite{yu2018magnon}, and microwave devices~\cite{tsai2005tunable}. Placing a YIG sphere within a cavity capitalizes on its unique properties, enabling the creation of a sensitive and easily tunable system~\cite{zhang2015magnon, wang2022dissipation, sohail2023entanglement, sohail2023distant, Sohail:23}. The YIG sphere functions as a mechanical generator, with its movement determined by its magnetization.

Researchers have demonstrated magnon-induced transparency (MIT) within a magnomechanical cavity system, establishing a connection between the movement of a YIG sphere and the flow of light within a cavity~\cite{zhang2016cavity}. Notably, researchers have achieved the cooling of a YIG sphere to its quantum ground state~\cite{asjad2023magnon}, and have established connections between the YIG sphere and the light inside the cavity~\cite{wu2021remote, zheng2021enhanced}. Recently, researchers have examined fano-type optical response and four-wave mixing studied in magnetoelastic system which has applications in highly sensitive detection and quantum information processing \cite{sohail2023controllable}.

Recently, researchers have noted the emergence of magnonic frequency combs in the context of optics. This intriguing phenomenon, detailed in an article referenced as \cite{articlecomb}, describes a spectrum characterized by discrete frequency components evenly spaced at regular intervals. Magnonic frequency combs have garnered attention due to their potential applications in diverse scientific disciplines. They play a significant role in enhancing the precision of atomic clocks, where the evenly spaced frequencies facilitate accurate timekeeping mechanisms. Recently some novel experimental research has demonstrated magnonic frequency combs\cite{xu2023magnonic}, slow-light hybird magnonics\cite{xu2024slow}, and magnonic switch\cite{wang2021inverse}.
These findings pave the way for innovative technologies, including highly sensitive magnetic sensors~\cite{xu2023optomechanical} and advancements in quantum information processing~\cite{sheng2023nonlocal}.

In this study, we propose a magnomechanical cavity system utilizing magnetic dipole interactions to achieve robust coupling between the collective motion of a large number of spins in a ferrimagnet. Magnomechanical cavities demonstrate exceptional characteristics, including strong coupling~\cite{flower2019experimental}, hybrid functionality~\cite{lachance2019hybrid}, high tunability, and potential applications in precision measurement, signal processing, and information storage.

Motivated by the intriguing possibilities \cite{li2018magnon}, we investigate cavity magnomechanics and analyze the behavior of a diffraction grating within a cavity magnon set up in the presence of a robust standing-wave field. In our proposed scheme strong standing wave pump and weak probe fields are optical and are applied from the left side of the microwave cavity while the magnon is driven by a weak biased microwave field that is directly applied on it in a perpendicular direction to generate phonon modes. Our study reveals a captivating relationship between the diffraction grating and the strength of coupling between the cavity magnon, denoted as $g_{\text{am}}$, and the phonon modes denoted as $g_{\text{mb}}$.

The cavity magnomechanics system provides a versatile means to control the diffraction grating through the coupling strengths $g_{\text{mb}}$ and $g_{\text{am}}$ which facilitate the transfer and storage of energy to different orders of the diffraction grating. This capability presents a promising avenue for tuning the grating to meet specific application requirements. Our goal is to make use of the established mechanism by which magnons and thermal vibrations may couple to generate multiple MIT which further leads to MMIG.

The structure of the article is organized as follows: In Section~\ref{Sec:model}, we derive the quantum Langevin equations (QLEs) from the Heisenberg equation of motion. Employing a standard input-output connection, we establish a mathematical representation for the entering field. Then we calculate the periodic manipulation of the propagation characteristics of probe beam. Finally, we determine the Fraunhofer diffraction intensity and investigate the transmission of the probe field to various diffraction orders. Section~\ref{Sec:results} presents numerical findings of our proposed system and in Section~\ref{Sec:discussion}, discussions related to the MMIG. In the concluding Section~\ref{Sec:conclusions}, we summarize our work and highlight key insights gained from the study.
\begin{figure}
\centering
\includegraphics[width=0.9\linewidth]{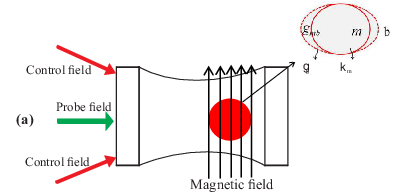}
\includegraphics[width=0.8\linewidth]{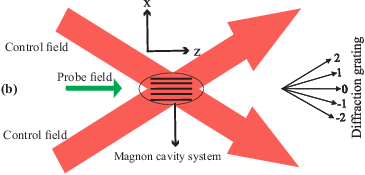}\par\medskip
\caption{Schematic of an optical system comprising cavity mode $a$ and magnon modes $m$. The cavity modes interact with magnon modes through coupling $g_{\text{am}}$ and with phonon modes via coefficient $g_{\text{mb}}$. The probe and control fields are applied on the left side of the cavity magnon system.
b) Schematics illustrating a probe field that diffracts via a position-dependent standing wave (SW) control field, whereby the entire magnon cavity system operates as an aperture, generating a diffraction grating.
The right side depicts the diffraction grating orders when the probe light beam undergoes diffraction.}
\label{figure1}
\end{figure}
\section{Theoretical Model}
\label{Sec:model}

The proposed model is illustrated in Fig.~\ref{figure1}. The cavity interior is driven by both the probe field and an intense pumping field originating from the left side of the cavity. A microwave source with frequency $\omega_\text{m}$ serves as an excitation for the magnons. Figure~\ref{figure1}(a) provides a schematic representation of a hybrid magnomechanical system, consisting of a Fabry-Perot cavity with length $L$ accommodating cavity photon and magnon modes. 
When a microwave field is directly applied to magnons, the mechanical vibrations induced by the magnons generate phonon modes. These magnon modes arise from the collective motion of numerous spins in a ferrimagnet, such as a YIG sphere (a sphere with a diameter of $250~\mu$m). 
The coupling between magnons and cavity photons is facilitated by a magnetic dipole interaction.

The connection between these magnons and phonons is established through magnetostrictive coupling. Specifically, a magnon excitation within a YIG sphere induces a changing magnetization that deforms the geometric shape within the sphere, resulting in the generation of vibration modes (phonons). Two strong pump beams are symmetrically shifted to the $z$ axis, as shown in Fig.~\ref{figure1}(b). They impact the cavity medium at angles that interact generating a standing wave (SW) within the medium with a spatial interval in the transversal $x$-direction.

We expect to see a periodic change in these coefficients as the SW adjusts across $x$ from nodes to antinodes due to the effect of the pump fields on the absorption as well as the dispersion of the weak probe field. The spatial periodic modulation of phase as well as amplitude, causes the weak probing field to diffract into various orders as it goes through the cavity magnon system. When the cavity magnon coupling is modified, the SW field generates this periodic modulation.

The system Hamiltonian is
\begin{align}
\label{eq:ham1}
H=&H_0+H_{\text{int}}+H_{\text{dr}},
\end{align}
where
\begin{align}
H_0=&\Delta_{\text{a}}a^{\dagger}a+\frac{\omega_{\text{b}}}{2}(x^{2}+p^{2})+\Delta_{\text{m}}m^{\dagger}m,\nonumber\\
H_{\text{int}}=&g_{\text{am}}(m^{\dagger}a+a^{\dagger}m)+g_{\text{mb}}m^{\dagger}mx,\nonumber\\
H_{\text{dr}}=&i(\mathcal{E}_{\text{m}}m^{\dagger}\text{e}^{-\text{i}\delta_{\text{ml}}t}-\mathcal{E}_{\text{m}}^{*}m \text{e}^{\text{i}\delta_{\text{ml}}t})+\text{i}(a^{\dagger}\mathcal{E}_{\text{p}}\text{e}^{-\text{i}\delta_{\text{pl}}t}\nonumber\\
&-a\mathcal{E}^{*}_{\text{p}}
\text{e}^{i\delta_{\text{pl}}t})+\text{i}(\mathcal{E}_{\text{l}}a^{\dagger}-\mathcal{E}^{*}_{\text{l}}a),\label{eq:ham2}
\end{align}
Here, $\Delta_{\text{a}}:=\omega_{\text{a}}-\omega_{\text{l}}$, $\Delta_{\text{m}}:=\omega_{\text{m}}-\omega_{\text{l}}$, $\delta_{\text{ml}}:=\Omega_{\text{m}}-\omega_{\text{l}}$ and $\delta_{\text{pl}}:=\omega_{\text{p}}-\omega_{\text{l}}$,
where $\omega_{\text{p}}$, $\omega_{\text{a}}$, $\omega_{\text{m}}$, $\omega_{\text{l}}$, $\Omega_{\text{m}}$ and $\omega_{\text{b}}$ represent the resonance frequencies of probe field, cavity modes, magnon modes, pump field, magnon driving field and phonon modes, respectively. Furthermore, $a^{\dagger}$ and $a$ are cavity mode creation and annihilation operators, whereas $m^{\dagger}$ and $m$ are magnon mode creation and annihilation operators. The dimensionless position and momentum quadratures of the mechanical mode are denoted by $x$ and $p$.

In Eq.~(\ref{eq:ham1}),
$H_0$ represents the free part of the Hamiltonian, $H_{\text{int}}$ denotes the interaction part, and $H_{\text{dr}}$ corresponds to the driving part of the Hamiltonian. In Eq.~\ref{eq:ham2}, the first part on the right side of $H_0$ provides the cavity mode annihilation and creation operators $a(a^{\dagger})$, the second part gives the mechanical mode operators, and the third term shows the magnon mode operators $m(m^{\dagger})$. The first part on the right side of $H_{\text{int}}$ in Eq.~\ref{eq:ham2} corresponds to the coupling of magnon modes with cavity modes with coupling strength $g_{\text{am}}$. The second term corresponds to the coupling of magnon modes with mechanical modes with coupling strength $g_{\text{mb}}$. The Hamiltonian $H_{\text{dr}}$ consists of the field directly applied to magnon modes, as well as the probe and pump fields applied from the left side of the magnon cavity. The amplitude of the probe and pump fields is
\begin{align}
\label{eq:probefield}
\mathcal{E}_{\text{p}}=&\sqrt{\frac{2\kappa_\text{a} P_\text{p}}{\hbar\omega_\text{p}}},
\end{align}
and 
\begin{align}
\label{eq:pumpfield}
\mathcal{E}_{\text{l}}=&\sqrt{\frac{2\kappa_{\text{a}} P_\text{d}}{\hbar \omega_\text{l}}},
\end{align}
respectively, where $\kappa_{\text{a}}$ denotes the cavity decay rate, $P_\text{p}$ ({$P_\text{d}$}) represents the power of the probe (pump) field, and $\omega_{\text{p},\text{l}}$ describes the frequency of the probe and pump fields. The  interaction of weak microwave source applied to magnon modes is \cite{li2018magnon, li2020phase}
\begin{align}
\label{eq:microwavefield}
\mathcal{E}_{\text{m}}=&\frac{\sqrt{5}}{4}\gamma\sqrt{N}B_0,
\end{align}
where $\gamma$ denotes the gyromagnetic ratio, $N$ is the total number of spins inside the YIG sphere, and $B_0$ is the amplitude of the driving magnetic field.

Now we'll look at the Heisenberg formulation that describes motion to determine the way it describes system dynamics. For every generic operator $\mathcal{O}$, an expression is
\begin{align}
\label{eq:heisen}
\frac{\text{d}\mathcal{O}}{\text{d}t}=&-\frac{\text{i}}{\hbar}[\mathcal{O},H]-\gamma \mathcal{O}+\mathcal{N},
\end{align}
where $\gamma$ indicates the decay rate linked with the cavity photon, magnon, and phonon modes, while $\mathcal{N}$ represents the Brownian along with input vacuum noise operator related to the cavity field. When we apply Eq.~(\ref{eq:ham2}) in Eq.~(\ref{eq:heisen}), we get the following coupled equations:
\begin{align}
\dot{a}=&-\left(\kappa_{\text{a}}+\text{i}\Delta_\text{a}\right)a-\text{i}g_{\text{am}}m+\mathcal{E}_{\text{p}}\text{e}^{-\text{i}\delta_\text{pl} t}+\mathcal{E}_{\text{l}}\nonumber\\+&\sqrt{2\kappa_{a}}a_{\text{in}},\\
\dot{m}=&-(\text{i}\Delta_{\text{m}}+\kappa_{\text{m}})m-\text{i}g_{\text{am}}a-\text{i}g_{\text{mb}}mx+\mathcal{E}_{\text{m}}\text{e}^{-\text{i}\delta_\text{ml} t}\nonumber\\+&\sqrt{2\kappa_{\text{m}}}m_{\text{in}},\\
\dot{p}=&-\omega_{\text{b}}x-\gamma_{\text{b}}p-g_{\text{mb}}m^{\dagger}m+\zeta,\\
\dot{x}=&\omega_{\text{b}}p,
\end{align}
$\kappa_\text{a}$, $\gamma_\text{b}$, and $\kappa_\text{m}$, are the decay rates whereas the quantum noise operators for the cavity, magnon, and phonon modes are $a_{\text{in}}$, $m_{\text{in}}$, and $\zeta$, respectively. It is worth noting that the mean values of quantum noise, Brownian noise, along the input operator are all zero~\cite{qu2013phonon}.

Further we consider a much weaker probe field $\mathcal{E}_{\text{p}}$ and microwave field $\mathcal{E}_{m}$ than the pump field $\mathcal{E}_{l}$ to facilitate the solution of the aforementioned nonlinear quantum Langevin equations (QLEs). Consequently, we are able to define each operator as the average of the mean value and first-order quantum fluctuation term, i.e., $a=a_{\text{s}}+\delta a$, $x=x_{\text{s}}+\delta x$, $p=p_{\text{s}}+\delta p$, and $m=m_{\text{s}}+\delta m$. The steady-state solution for the aforementioned equations may be attained via setting the time derivatives to zero, namely,
\begin{align}
 a_{\text{s}}=&\frac{\mathcal{E}_{\text{l}}(\text{i}\Delta^{\prime}_{\text{m}}+\kappa_{\text{m}})}{(i\Delta^{\prime}_{\text{m}}+\kappa_{\text{m}})(\text{i}\Delta_{\text{a}}+\kappa_{\text{a}})+g^{2}_{\text{am}}},\\
 m_{\text{s}}=&\frac{-\text{i}g_{\text{am}}a_{\text{s}}}{i\Delta^{\prime}_{\text{m}}+\kappa_{\text{m}}},\\
 x_{\text{s}}=&\frac{-g_{\text{mb}}|m_{\text{s}}|^{2}}{\omega_{\text{b}}},
\end{align}
where $\Delta_\text{m}^\prime=\Delta_\text{m}-g_{\text{mb}}x_{\text{s}}$ is the effective detuning values of the magnon modes. The linearize QLEs of motion are expressed as follows:
\begin{align}
\delta\dot{a}=&-\left(\kappa_{\text{a}}+\text{i}\Delta_\text{a}\right)\delta a-\text{i}g_{\text{am}}\delta m+\mathcal{E}_{\text{p}}\text{e}^{-\text{i}\delta_{\text{pl}} t},\\
\delta\dot{m}=&-(\text{i}\Delta_{\text{m}}+\kappa_{\text{m}})\delta m-\text{i}g_{\text{am}}\delta a \nonumber\\-&\text{i}g_{\text{mb}}\left(m_{\text{s}}\delta x+x_{\text{s}}\delta m\right)+\mathcal{E}_{\text{m}}e^{-\text{i}\delta_{\text{ml}} t},\\
\delta\dot{p}=&-\omega_{\text{b}}\delta x-\gamma_{\text{b}}\delta p-g_{\text{mb}}\left(m^{*}_{\text{s}}\delta m+m_{\text{s}}\delta m^{\dagger}\right),\\
\delta\dot{x}=&\omega_{\text{b}}\delta p.
\end{align}
The linearized equations of motion are then solved perturbatively using ansatzes~\cite{boyd2010} $\delta\mathcal{O}=\sum_{n\rightarrow{\{-,+\}}}\mathcal{O}_{n}\text{e}^{\text{i}n\delta t}$, where $\mathcal{O}={a,m,x,p}$, with $\delta=\delta_{pl}=\delta_{ml}$. We consider the magnon driving field to become resonant with the probe field frequency which leads us to consider $\delta=\delta_{\text{pl}}=\delta_{\text{ml}}$ \cite{li2020phase}. In our suggested model, the oscillation in the cavity is predominantly due to magnomechanical phenomena caused by a directly applied magnetic field. Moreover, the amplitude and phase oscillations can be controlled by the control and probe lasers. This oscillation, in turn, induces Stokes and anti-Stokes dispersion in the control field. The first-order solution for the transmitted probe field is then obtained using the aforementioned methods.
\begin{align}
a_{-}=&\frac{\mathcal{M}}{\mathcal{R}},
\end{align}
where
\begin{widetext}
\begin{align}
\mathcal{M}=&\mathcal{E}_{\text{p}}(-\delta \alpha _5 \omega _\text{b}^2 g_{\text{mb}}^4 m_s^4-(-\alpha _1 \alpha _6-\text{i}\omega _\text{b} g_{\text{mb}}^2 m_\text{s} m^{*}_s)(\alpha _2\delta g_{\text{am}}^2+\alpha _3 (-\delta\omega_\text{b} g_{\text{mb}}^2 m_s m^{*}_s+\text{i}\alpha _2\alpha_4\delta)))\nonumber\\&-\text{i}\alpha _1 g_{\text{am}}\mathcal{E}_{m}(\alpha _2\delta  g_{\text{am}}^2+\alpha _3(-\delta\omega_\text{b}g_{\text{mb}}^2m_\text{s}m^{*}_\text{s}+\text{i}\alpha_2\alpha _4\delta)),\\
\mathcal{R}=&\alpha _1g_{\text{am}}^2(\alpha_2\delta g_{\text{am}}^2+\alpha _3(-\delta\omega_\text{b} g_{\text{mb}}^2 m_\text{s} m^{*}_s+\text{i}\alpha _2\alpha _4\delta))+\alpha _7 (-\delta \alpha _5\omega _b^2 g_{\text{mb}}^4 m_\text{s}^4-\left(-\alpha_1\alpha _6-\text{i}\omega_\text{b}g_{\text{mb}}^2m_\text{s} m^{*}_\text{s}\right)(\alpha _2\delta g_{\text{am}}^2\nonumber\\&+\alpha _3(-\delta\omega_\text{b}g_{\text{mb}}^2m_\text{s} m^{*}_\text{s}+\text{i}\alpha_2\alpha_4\delta))),\\
\end{align}
\end{widetext}
and
\begin{align}
\alpha_1=&-\omega_{\text{b}}^2+\delta(\delta+\text{i}\gamma_{\text{b}}),\\
\alpha_2=&\delta(\gamma_{\text{b}}+i\delta)-\text{i}\omega_{\text{b}}^2,\\
\alpha _3=&k_a+\text{i}(\delta+\Delta_\text{a}),\\
\alpha_4=&\delta+\Delta_{\text{m}}+g_{\text{mb}}x_{\text{s}}+\text{i}k_\text{m},\\
\alpha_5=&-\text{i}k_a+\delta +\Delta_\text{a},\\
\alpha_6=&k_m+\text{i}\left(-\delta +\Delta_{\text{m}}+g_{\text{mb}}x_{\text{s}}\right),\\
\alpha_7=&k_a-\text{i}(\delta-\Delta_{\text{a}}).
\end{align}
The input-output relationship can be expressed as \cite{Walls2007}
\begin{align}
\mathcal{E}_{\text{out}}(t)+\mathcal{E}_{\text{p}}\text{e}^{-\text{i}\delta t}+\mathcal{E}_{\text{l}}=&\sqrt{2\kappa_{\text{a}}}a,\label{eout}
\end{align}
where
\begin{align}
\mathcal{E}_{\text{out}}(t)=&\mathcal{E}^0_{\text{out}}+\mathcal{E}^+_{\text{out}}\mathcal{E}_{\text{p}} \text{e}^{-i\delta t}+\mathcal{E}^-_{\text{out}}\mathcal{E}_{\text{p}}\text{e}^{\text{i}\delta t}.\label{eout2}
\end{align}
After solving Eqs. (\ref{eout}) and (\ref{eout2}), we obtain 
\begin{align}
\mathcal{E}^+_{\text{out}}=&\frac{\sqrt{2\kappa_{a}}a_-}{\mathcal{E}_{\text{p}}}-1.
\end{align}
The homodyne method could be used to measure it \cite{Walls2007}. For the sake of simplicity, we define
\begin{align}
\mathcal{E}^+_{\text{out}}+1=\frac{\sqrt{2\kappa_{\text{a}}}a_-}{\mathcal{E}_{\text{p}}}=&\mathcal{E}_{\text{T}}.
\end{align}
The mathematical quadrature formula for the field $\mathcal{E}_{\text{T}}$ is written as
\begin{equation}
\mathcal{E}_{\text{T}}=\text{Re}[\mathcal{E}_{\text{T}}]+\text{i}\text{Im}[\mathcal{E}_{\text{T}}].
\end{equation}
The out-of-phase along with in-phase quadratures for the resulting probe field is represented by $\text{Re}[\mathcal{E}_{\text{T}}]$ and $\text{Im}[\mathcal{E}_{\text{T}}]$, respectively.
\subsection{Dynamics of MMIG}
Using Maxwell's equation, the wave propagation caused by the probing light beam ($\mathcal{E}_\text{p}$) may be \cite{kuang2011gain}
\begin{align}
\frac{\text{d}\mathcal{E}_{\text{p}}}{\text{d}z}=&[-\eta(\text{x})+\text{i}\zeta(\text{x})]\mathcal{E}_{\text{p}},\label{pro-eq}
\end{align}
in which $\eta(\text{x})=(\frac{2\pi}{\lambda})\text{Re}[\mathcal{E}_{\text{T}}]$ as well as $\zeta(\text{x})=(\frac{2\pi}{\lambda})\text{Im}[\mathcal{E}_{\text{T}}]$ denote absorption as well as dispersion corresponding to the probing field with a wavelength $\lambda$.  

To simplify the analysis, we focus on the MMIG characteristics, disregarding the transversal component of Eq.~(\ref{pro-eq}) \cite{kuang2011gain}. This equation can be employed to straightforwardly calculate the optical transmission function describing the probe laser beam at $z=L$, given by

\begin{align}
T_{\text{trans}}(x)=&e^{-\eta(x)L+\text{i}\zeta(x)L},
\label{transmission}
\end{align}

where $|T_{\text{trans}}(x)|=\text{e}^{\eta(x)L}$ and $\phi(x)=\zeta(x)L$ denote the magnitude and phase modulation within the cavity, respectively. Here it is to be mentioned that the standing-wave control field generates a periodic modulation in the system, leading to spatial variations in the transmission function. As a result, the transmission of the system will exhibit different behaviors at different positions along the x-direction as shown in Fig.\ref{figure1}(b).
The intensity distribution of diffraction may be shown by considering the incident probing field as a plane wave \cite{kuang2011gain}
\begin{align}
I(\theta)=&{\vert E(\theta)\vert}^{2}\times\frac{\sin^{2}(N \pi \Lambda_x\sin\theta/\lambda)}{N^{2}\sin^{2}(\pi \Lambda_x\sin\theta/\lambda)},
\label{diff-int}
\end{align}
where $\theta$ is the angle of diffraction in the $x$-direction, $\Lambda_{x}=\nicefrac{\pi}{k_{x}}$ is the spatial period within the $x$ direction, and $N$ is a particular quantity of spatial periods generating an atomic grating. The term $E(\theta)$ denotes the Fourier transformation of $T_{\text{trans}}(x)$ and describes a Fraunhofer diffraction over just one period given by
\begin{align}
E(\theta)=&\int_0^{1} T_{\text{trans}}(x)e^{-2\pi\text{i} \Lambda_x x \sin\theta/\lambda}\text{d}x,
\end{align}
$\theta$ represents the diffraction angles with respect to the z-direction.
\begin{figure*}
\centering
\includegraphics[width=0.5\linewidth]{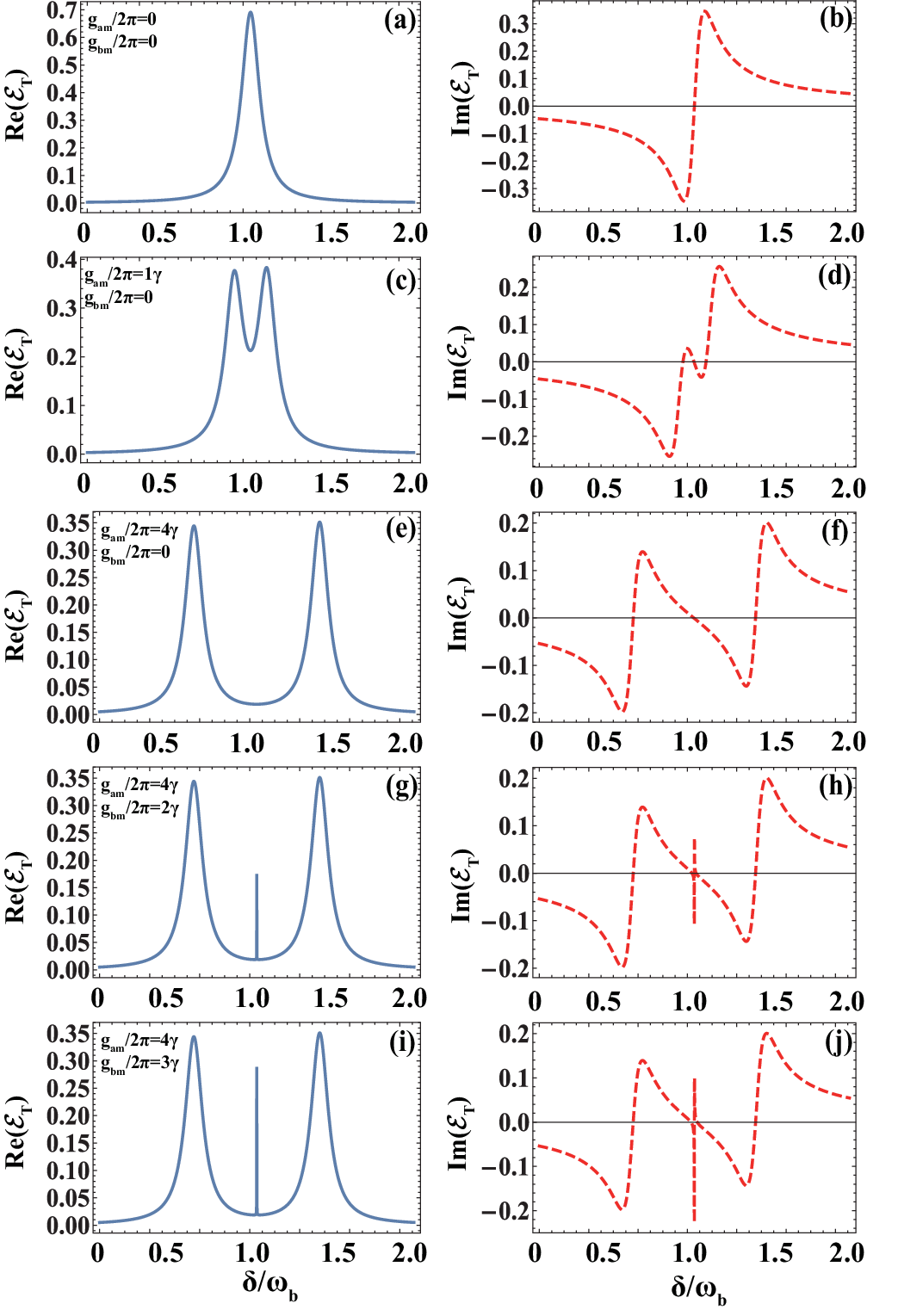}
\caption{The absorption and dispersion parts of output probe field versus probe field detuning $\delta$. The fixed parameters considered are $\omega_{\text{b}}/2\pi=10\gamma$, $\kappa_{\text{a}}=\omega_{\text{b}}/15$, $\kappa_{\text{m}}=\omega_{\text{b}}/15$, $\Delta_{\text{m}}/2\pi=10\gamma$, $\Delta_{\text{a}}/2\pi=10\gamma$, $\gamma_{\text{b}}/2\pi=0.0014\gamma$, $\gamma=1$MHz.}
\label{absorption-dispersion}
\end{figure*}

\begin{figure*}
\centering
\includegraphics[width=0.6\linewidth]{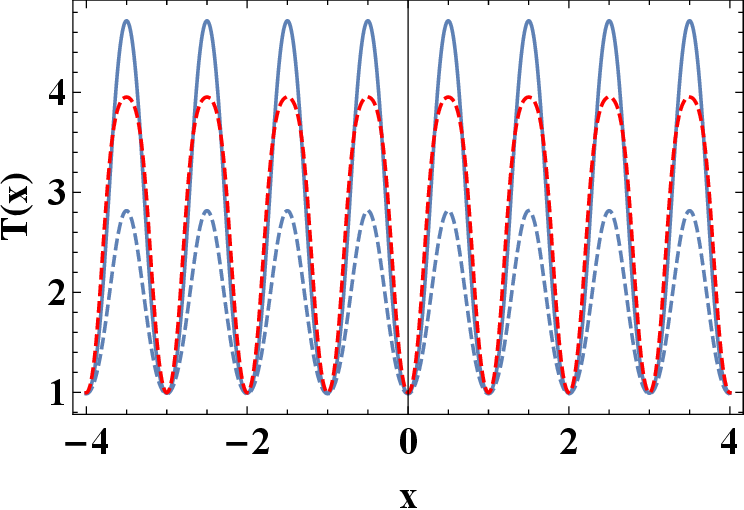}
\caption{The normalized transmission profile as a function of position $x$ with the variation of cavity magnon coupling $g_{\text{am}}$, i.e. (a) $g_{\text{am}}/2\pi=1\gamma$ (blue dashed line), (b) $g_{\text{am}}/2\pi=2\gamma$ (red dashed line), (c) $g_{\text{am}}/2\pi=4\gamma$ (blue solid line). The fixed parameters considered are $\omega_{\text{b}}/2\pi=10\gamma$, $\kappa_{\text{a}}=\omega_{\text{b}}/15$, $\kappa_{\text{m}}=\omega_{\text{b}}/15$, $\Delta_{\text{m}}/2\pi=10\gamma$, $\Delta_{\text{a}}/2\pi=10\gamma$, $\gamma_{\text{b}}/2\pi=0.0014\gamma$, $\gamma=1$MHz.}
\label{transmission1}
\end{figure*}
\begin{figure*}
\centering
\includegraphics[width=0.6\linewidth]{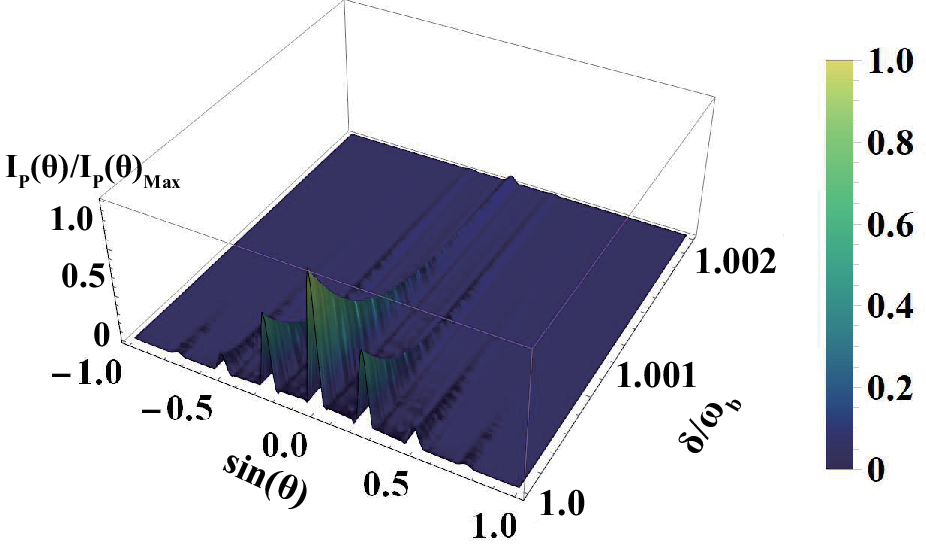}
\caption{The intensity of normalized diffraction against $sin(\theta)$ and detuning $\delta$. The fixed parameters considered are $g_{\text{mb}}/2\pi=3\gamma$, $g_{\text{am}}/=3\gamma$, $\omega_{\text{b}}/2\pi=10\gamma$, $\kappa_{\text{a}}=\omega_{\text{b}}/15$, $\kappa_{\text{m}}=\omega_{\text{b}}/15$, $\Delta_{\text{m}}/2\pi=10\gamma$, $\Delta_{\text{a}}/2\pi=10\gamma$, $\gamma_{\text{b}}/2\pi=0.0014\gamma$, $\gamma=1$MHz, $\mathcal{E}_{m}=0.001\gamma$, interaction length $L=55$mm.}
\label{3Ddiffraction}
\end{figure*}
\begin{figure*}
\centering
\includegraphics[width=0.8\linewidth]{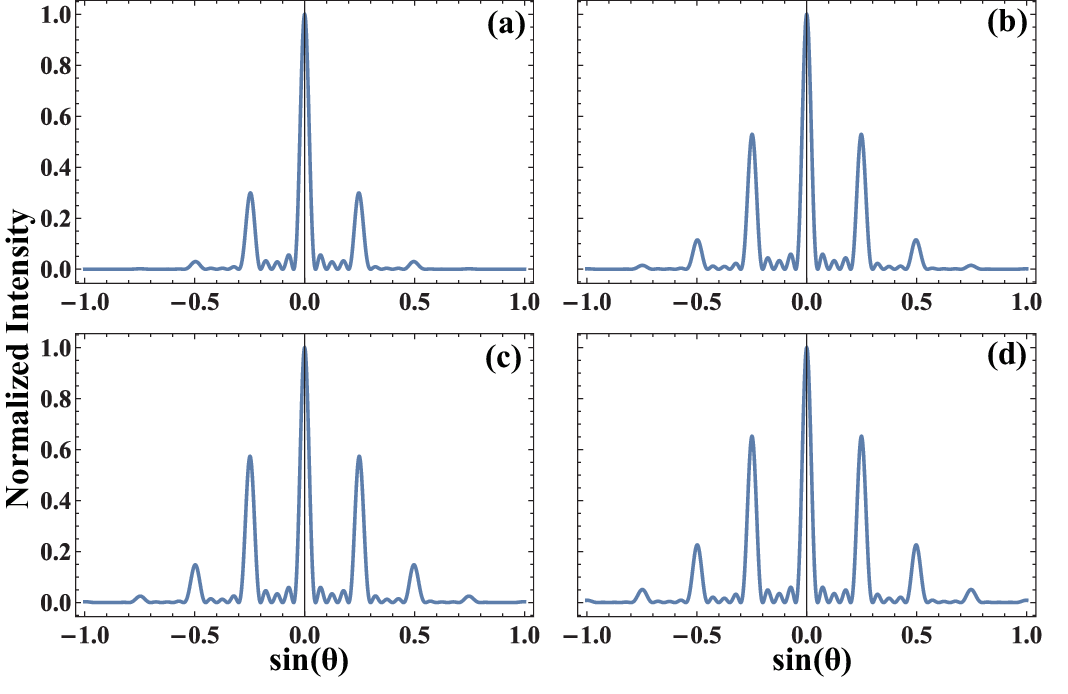}
\caption{The intensity of normalized diffraction as a function of $\text{sin}(\theta)$ with the variation of cavity magnon interaction $g_{\text{am}}$ and magnon phonon interaction $g_{\text{am}}$. (a) $g_{\text{am}}/2\pi=1\gamma$, (b) $g_{\text{am}}/2\pi=2\gamma$, (c) $g_{\text{am}}/2\pi=3\gamma$, (d) $g_{\text{am}}/2\pi=4\gamma$. The fixed parameters considered are $g_{\text{mb}}/2\pi=3\gamma$, $\omega_{\text{b}}/2\pi=10\gamma$, $\kappa_{\text{a}}=\omega_{\text{b}}/15$, $\kappa_{\text{m}}=\omega_{\text{b}}/15$, $\Delta_{\text{m}}/2\pi=10\gamma$, $\Delta_{\text{a}}/2\pi=10\gamma$, $\gamma_{\text{b}}/2\pi=0.0014\gamma$, $\mathcal{E}_{m}=0.001\gamma$, $|\mathcal{E}_{\text{l}}/2\pi|=4.5\gamma$, $\gamma=1$MHz.}
\label{grating1}
\end{figure*}
\begin{figure*}
\centering
\includegraphics[width=0.8\linewidth]{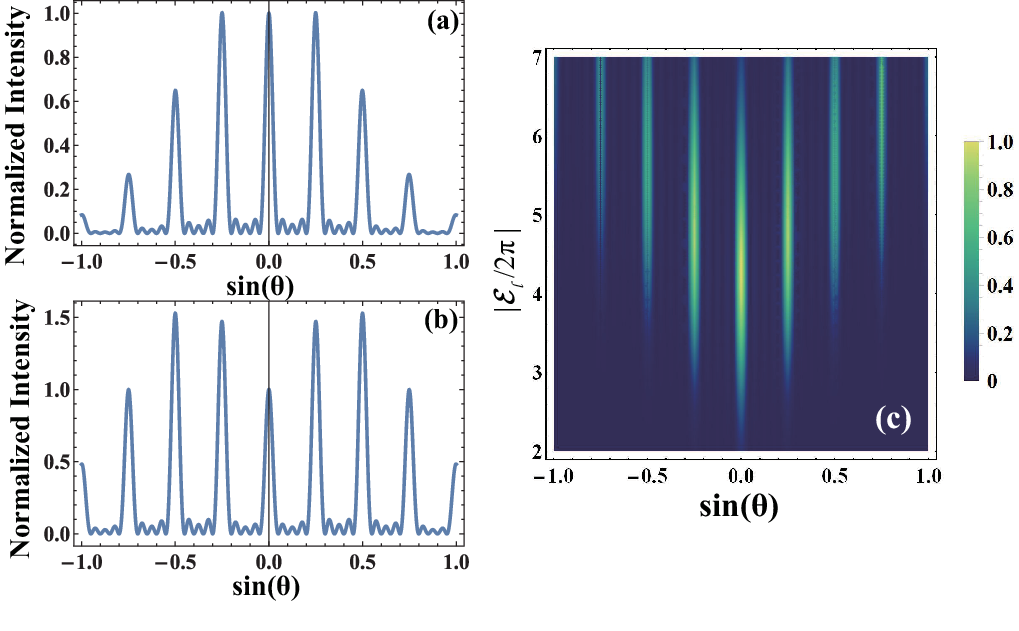}
\caption{The intensity of normalized diffraction as a function of $\text{sin}(\theta)$ with the variation of pump field. (a) $|\mathcal{E}_{\text{l}}/2\pi|=5\gamma$, (b) $|\mathcal{E}_{\text{l}}/2\pi|=6\gamma$, (c) density plot of normalized diffraction intensity versus $sin(\theta)$ and the intensity of pump field $|\mathcal{E}_{\text{l}}/2\pi|$. The fixed parameters considered are $g_{\text{am}}/2\pi=4\gamma$, $g_{\text{mb}}/2\pi=3\gamma$, $\omega_{\text{b}}/2\pi=10\gamma$, $\kappa_{\text{a}}=\omega_{\text{b}}/15$, $\kappa_{\text{m}}=\omega_{\text{b}}/15$, $\Delta_{\text{m}}/2\pi=10\gamma$, $\mathcal{E}_{m}=0.001\gamma$, $\Delta_{\text{a}}/2\pi=10\gamma$, $\gamma_{\text{b}}/2\pi=0.0014\gamma$, $\gamma=1$MHz.}
\label{grating2}
\end{figure*}
\section{Results}
\label{Sec:results}
Our research on MMIG with a standing wave field pump in a cavity is presented in this section. We investigated the effects of various system parameters on phase as well as amplitude modulation, which impacts the controllability of the diffraction patterns along with intensity along with the transmission profile of the probing light beams. We analyze the intensity of MMIG by varying the coupling strength $g_{\text{am}}$ between magnon and cavity. We also investigate the effect of the coupling strength $g_{\text{mb}}$ between the phonon along with magnon. Furthermore, we analyze the effect of the interaction length on MMIG intensities. The fixed parameters considered are $\omega_{\text{b}}/2\pi=10\gamma$, $\kappa_{\text{a}}=\omega_{\text{b}}/15$, $\kappa_{\text{m}}=\omega_{\text{b}}/15$, $\Delta_{\text{m}}/2\pi=10\gamma$, $\Delta_{\text{a}}/2\pi=10\gamma$, $\gamma_{\text{b}}/2\pi=0.0014\gamma$, $\gamma=1$MHz.

To begin, we'll look at how the output probe light behaves with or without cavity modes, as well as the magnon mode interaction strength $g_{\text{am}}$. In Fig.~\ref{absorption-dispersion}, we present the absorption ($\text{Re}[\mathcal{E}_{T}]$) and dispersion ($\text{Im}[\mathcal{E}_{T}]$) of the probe light beam in the cavity magnon system against a normalized detuning $\delta/\omega_{b}$ for various values of $g_{\text{am}}$ and $g_{\text{mb}}$. The $\text{Im}[\mathcal{E}_{T}]$ of the output probe field illustrates the dispersive characteristics of the magnomechanical cavity for specific $g{\text{am}}$ and $g_{\text{mb}}$. The slope can change with varying values of $g_{\text{am}}$ and $g_{\text{mb}}$. Specifically, a negative slope corresponds to a negative group index, and a positive slope leads to a positive group index of the cavity. However, our main focus is on the transmission of the probe field of the output spectrum $\text{Re}[\mathcal{E}_{\text{T}}]$.

In the scenario where $g_{\text{am}}$ and $g_{\text{mb}}$ are absent, the real and imaginary components of the output probe field are presented against the probe detuning $\delta/\omega_{b}$ in Fig.~\ref{absorption-dispersion}(a) and (b). Illustrated by the solid blue curve, the graph exhibits a Lorentzian profile, indicating substantial absorption of the probe light within the cavity (refer to Fig.~\ref{absorption-dispersion}(a)). It is notable that all incident light is absorbed within the magnon cavity system. In the absence of magnon-phonon interaction ($g_{\text{mb}}=0$) and with a fixed cavity-magnon coupling of $g_{\text{am}}/2\pi=1$ MHz, a narrow transmission window appears in the probe field spectrum, depicted by the blue curve. This narrow transparency window corresponds to a minimal transparency region with associated dispersion changes, as illustrated in Fig.~\ref{absorption-dispersion}(c) and (d).
Upon increasing the cavity-magnon coupling to $g_{\text{am}}/2\pi=4$ MHz, a broader transparency window with anomalous dispersion emerges, as shown in Fig.~\ref{absorption-dispersion}(e) and (f).
Introducing a small increment in magnon-phonon interaction strength ($g_{\text{mb}}/2\pi=2$ MHz) while maintaining the cavity-magnon coupling at $g_{\text{am}}/2\pi=4$ MHz, the transparency window starts to bifurcate into a double window with the associated dispersion profile depicted in Fig.~\ref{absorption-dispersion}(g) and (h).
Further increasing the magnon-phonon interaction to $g_{\text{mb}}/2\pi=3$ MHz, while keeping the cavity-magnon coupling constant at $g_{\text{am}}/2\pi=4$ MHz, results in a more pronounced splitting of the transparency window into a double Magnon-Mechanically Induced Transparency (MMIT) scenario, as shown in Fig.~\ref{absorption-dispersion}(i) and (j).

\begin{figure*}
\centering
\includegraphics[width=0.8\linewidth]{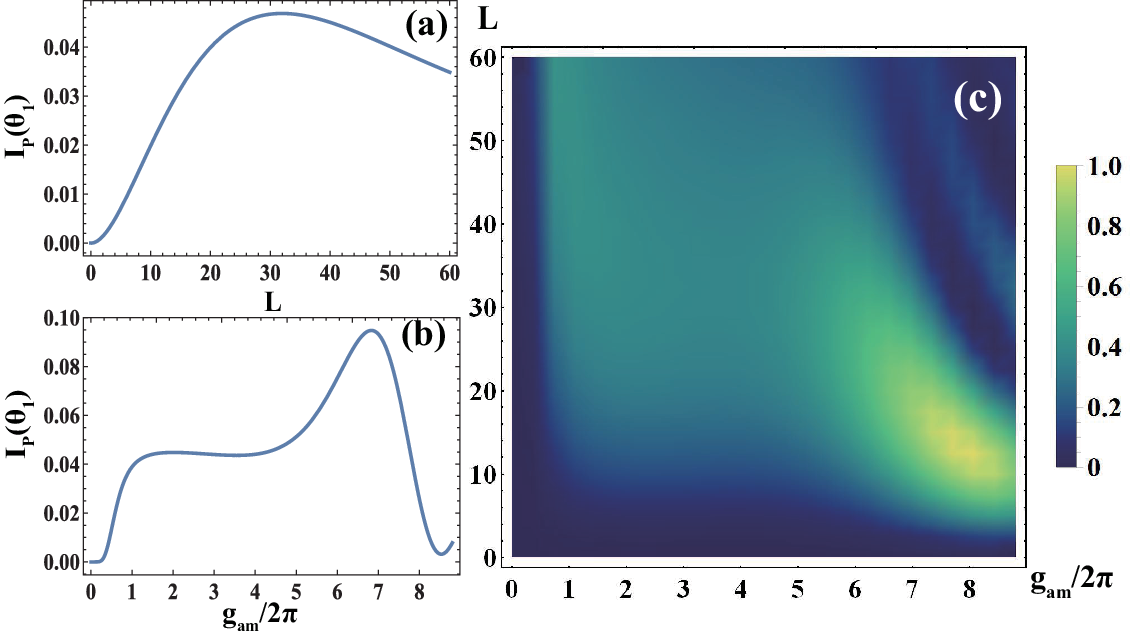}
\caption{The intensity of first order diffraction. (a) Fixed coupling strength $g_{\text{am}}/2\pi=4\gamma$, (b) Fixed interaction length $\text{L}=20$mm, (c) Density plot of normalized first order diffraction intensity against cavity magnon coupling strength $g_{\text{am}}/2\pi$ and interaction length L. The fixed parameters considered are $g_{\text{am}}/2\pi=4\gamma$, $g_{\text{mb}}/2\pi=3\gamma$, $\omega_{\text{b}}/2\pi=10\gamma$, $\kappa_{\text{a}}=\omega_{\text{b}}/15$, $\kappa_{\text{m}}=\omega_{\text{b}}/15$, $\Delta_{\text{m}}/2\pi=10\gamma$, $\mathcal{E}_{m}=0.001\gamma$ $\Delta_{\text{a}}/2\pi=10\gamma$, $\gamma_{\text{b}}/2\pi=0.0014\gamma$, $\gamma=1$MHz.}
\label{first order}
\end{figure*}

Now, let us delve into the modulation of the probe light beam transmission, a key aspect in achieving the desired diffraction intensity pattern.
In this analysis, we consider the pump field as a SW field, represented by
\begin{equation}
\mathcal{E}{_\text{l}}=\sqrt{\frac{2\kappa\text{a} P_\text{d}}{\hbar \omega_\text{l}}}\sin\left[\pi \text{x}/\Lambda_{\text{x}}\right],
\end{equation}
where $\Lambda_{\text{x}}$ is the spatial period. Other parameters are kept consistent with those in Fig.~\ref{absorption-dispersion}. The SW control field plays a crucial role in creating MMIT.
Due to the intensity-dependent response in the cavity, the SW control field induces spatially modulated absorption and refraction of the probe field. Consequently, the entire cavity acts as a grating, allowing the probe beam to diffract in various directions. Fig.~\ref{transmission1} illustrates the transmitted probe beam against the position $x$ for different values of cavity magnon coupling. Recalling our earlier observation from Fig.~\ref{absorption-dispersion} that weak magnon-cavity coupling results in enhanced absorption, increasing the coupling strength $g_{\text{am}}$ leads to MMIT, resulting in reduced absorption and enhanced transmission of the probe light beam. Let us consider $g_{\text{am}}/2\pi=1\gamma$ as an example, and observe the effect on the transmitted probe light in Fig.~\ref{transmission1} (depicted by the blue dashed line). The periodic modulation in the transmitted probe light beam is a direct consequence of the standing wave pump field.

At transverse locations of the SW field, specifically at nodes, the impact of the pump field strength is notably weak, resulting in insufficient modulation of the probe beam. Consequently, there is minimal transmission observed in the magnon cavity, leading to a lower overall amplitude. In contrast, at antinodes, the coupling is robust, significantly enhancing the transmission of the probe light beam. Despite this improvement, the overall amplitude remains subdued due to the relatively weak magnon cavity coupling (see Fig.~\ref{transmission1}, blue dashed line). Upon a slight increment in magnon cavity coupling ($g_{\text{am}}/2\pi=2\gamma$), there is a further boost in the transmission profile of the probe light beam. This enhancement can be attributed to reduced absorption, leading to an augmented MMIT effect (see Fig.~\ref{transmission1}, red dashed line). With a magnon cavity coupling of $g_{\text{am}}/2\pi=4\gamma$, a broader transparency window emerges, indicative of a strong magnon cavity interaction. This results in a significantly enhanced transmission profile of the probe light beam (see Fig.~\ref{transmission1}, blue solid line).

Furthermore, we will investigate the interference pattern exhibited by the probe light that traverses the SW pump field, acting as a slit. In addition, by adjusting the cavity magnon interaction parameter $g_{\text{am}}$, we will investigate the diffraction intensity pattern orders in the far-field regime, which is also referred to as Fraunhofer diffraction. In Fig.~\ref{3Ddiffraction}, we depict the normalized diffraction intensity plotted against $\sin\theta$ and detuning $\delta$. The figure clearly illustrates a rapid decrease in diffraction intensity when the detuning slightly deviates from the transparency window, particularly for $\delta=\omega_{\text{b}}$. Consequently, for the subsequent discussion, we opt for a detuning $\delta$ equal to $\omega_{\text{b}}$.

Examining Fig.~\ref{grating1}(a) for a cavity magnon interaction of $g_{\text{am}}/2\pi=1\gamma$, we observe that the predominant energy of the probe field is concentrated at the central maximum. Additionally, there is only a relatively small diffraction of probe energy into higher diffraction orders. At $g_{\text{am}}/2\pi=1\gamma$, the weak transmitted amplitude grating is a consequence of probe field absorption within the cavity, resulting in the loss of higher orders of diffraction. As seen in Fig.~\ref{grating1}(b) for $g_{\text{am}}/2\pi=2\gamma$, the rate of probe transfer of energy increases gradually from zeroth through higher diffraction orders. This behavior is attributed to the heightened transmission of the probe light beam with the increase in $g_{\text{am}}/2\pi$, leading to amplified diffraction of light into higher orders. Similarly, in Fig.~\ref{grating1}(c) at $g_{\text{am}}/2\pi=3\gamma$, the enhancement in diffraction intensity from the zeroth to the first order is notable due to a further increase in the grating amplitude of the transmitted probe beam. According to Figure~\ref{transmission1} (solid blue line), the second order of diffraction intensity occurs at $g_{\text{am}}/2\pi=4\gamma$, where the grating amplitude is at its maximum, as seen in Figure~\ref{grating1}(d). It is important to note that a strong cavity magnon interaction, as depicted in Fig.~\ref{transmission1} (solid blue line), can result in significant non-zero phase modulation. This modulation could diffract a portion of the probe energy toward higher-order diffraction.

To achieve a more pronounced enhancement in the transfer of probe energy from the zeroth order to higher-order diffraction intensities, we escalate the amplitude of the SW pump field to $|\mathcal{E}_{\text{l}}/2\pi|=5\gamma$. This results in a further augmentation of the transmission grating profile, leading to increased transfer of probe energy to higher diffraction patterns, as depicted in Fig.~\ref{grating2}(a). Continuing the increase in the strength of the pump beam to $|\mathcal{E}_{\text{l}}/2\pi|=6\gamma$, we observe a decrease in the amplitude of the zeroth order, with probe energy now transferring from the zeroth order to the first, second, and third order diffraction orders, as illustrated in Fig.~\ref{grating2}(b). To visualize this enhancement in the transfer of probe energy to higher diffraction orders more clearly, a density plot in Fig.~\ref{grating2}(c) present the diffraction intensity against $\sin\theta$ and $|\mathcal{E}_{\text{l}}/2\pi|$. As the strength of the pump field increases, a discernible transfer of probing energy to higher orders becomes evident.

To investigate the influence of interaction length on the transfer of energy to the first-order diffraction pattern, we plot the diffraction intensity of the first order, designated as $I_{\text{p}}(\theta_{1})$, versus the interaction length $L$ in Fig.~\ref{first order} (a). The plot reveals that the magnitude of the first-order diffraction intensity increases with an increase in the interaction length within a certain range ($\text{L}\leq30$). However, it experiences a slight decrease with a further increase in the interaction length, as illustrated in Fig.~\ref{first order}(a).
Moving on to the examination of the effect of magnon cavity coupling strength $g_{\text{am}}$ on the first-order diffraction intensity, Fig.~\ref{first order}(b) depicts the first-order intensity of diffracted light versus cavity magnon coupling $g_{\text{am}}$. It is observed that for small values of $g_{\text{am}}$, the amplitude of the first-order diffraction intensity remains small. However, with a subsequent increase in $g_{\text{am}}$, the amplitude of the diffraction intensity shows an increment within a certain range. This behavior is attributed to the fact that when the cavity magnon interaction is weak, more probe light is absorbed inside the cavity, resulting in lower transmission of the probe light beam. In such a scenario, the transfer of probe energy to the first-order diffraction intensity is weak, leading to a lower amplitude.
Further analysis reveals that there is a specific cavity magnon interaction strength $g_{\text{am}}$ at which maximum diffraction occurs, accompanied by the maximum transfer of probe energy to the first-order diffraction intensity. This optimal value of $g_{\text{am}}$ leads to the development of the highest amplitude. Finally in Fig.~\ref{first order}(c), we present a density plot suggesting the existence of an optimal point for maximum energy transfer, influenced by both the coupling strength $g_{\text{am}}$ and the interaction length $L$.

\section{Discussion}
\label{Sec:discussion}
We provide a quick summary of our investigation and outcomes in this section. We have developed a method to achieve magnomechanically induced grating in a magnomechanical system as a result of our investigation and analysis. Three main parts make up this system: phonons, cavity microwave photons, and magnons. A phonon is a quantized vibration generated by directly applying a magnetic field to a magnon, while a cavity microwave photon is an electromagnetic wave that is quantized inside a resonant cavity. Magnons are quantized collective excitations of magnetic moments. By harnessing the interactions between these components, we have successfully generated a grating structure.

First, we establish a magnomechanically induced transparency by taking into account the Hamiltonian of the whole system along with driving the general coupled equation using the Heisenberg operator approach, which reveals the dynamics of each degree of freedom separately. Further, by using cavity input-output theory, from coupled Langevin equations, we extract the cavity output field, which indeed will contain the information of all optical interactions happening inside the cavity. The real (imaginary) part gives the absorption (dispersion) behavior of the output probe field. By appropriately choosing the quantum parameters such as cavity magnon interaction strength $g_{\text{am}}$ and magnon phonon interaction $g_{\text{mb}}$ the absorptive and dispersive properties of the medium are modified.

Together with Fraunhofer intensity diffraction calculations, we also employ SW control field to solve the probe light transmission function equations for MMIG. The transmission function varies spatially as a result of a periodic modulation in the system caused by the SW control field. Moreover, we are examining how the duration of medium interactions affects the pattern of diffraction intensities in MMIG. It is shown that most of the probing energies accumulates in the center when the interaction duration reaches a specific amount. As the interaction duration increased, the probe energy transferred to higher orders. Furthermore, the diffraction intensities of higher order may be attained by selecting the cavity magnon interacting strength $g_{\text{am}}$ in a suitable manner. 

These findings suggest that MMIG are crucial for generating high-efficiency gratings by considering the quantum parameters. As a result, properly adjusting the quantum parameters is a good way to achieve high diffraction efficiencies. Our approach offers a novel and promising method for achieving MMIG, which can have potential applications in information storage and retrieval. This advancement could contribute to the development of more efficient and versatile quantum memory systems, with implications for various fields such as quantum computing, quantum communication, and quantum information processing.
\begin{figure*}
    \centering
    \includegraphics[width=0.8\linewidth]{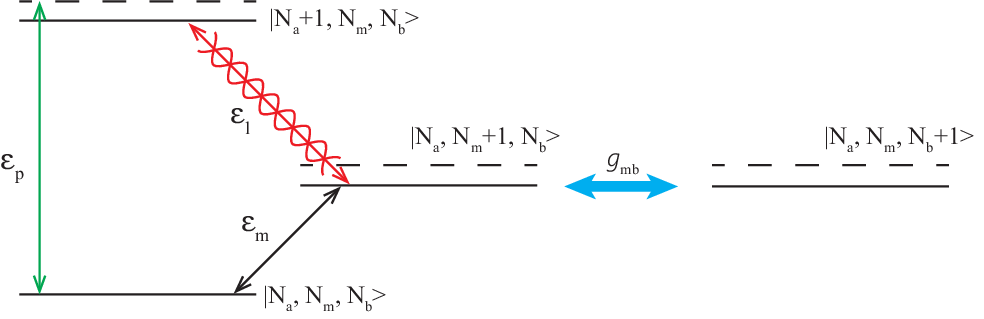}
    \caption{Energy level diagram. $\Ket{N_{a}}, \Ket{N_{m}}, \Ket{N_{b}}$ represents the states of photon, magnon, and phonon in microwave radiation.}
    \label{energylevel}
\end{figure*}
\section{Energy level diagram}
The energy level diagram in Fig.~\ref{energylevel} provides an understanding of the physical process behind this phenomenon. When there is no magnon-phonon interaction ($g_{\text{mb}}=0$), a photon is released during the probe photon transition, but the magnon state does not change $\Ket{N_{a}, N_{m}+1, N_{b}} \rightarrow \Ket{N_{a}+1, N_{m}, N_{b}} \rightarrow \Ket{N_{a}, N_{m}, N_{b}}$. The transition when a magnon is activated but the photon state stays unaltered is overlapped by $\Ket{N_{a}, N_{m}, N_{b}} \rightarrow \Ket{N_{a}+1, N_{m}, N_{b}} \rightarrow \Ket{N_{a}, N_{m}, N_{b}}$. Both the significant standing wave pump field as well as the cavity decay which generates this overlap lead to destructive interference. Consequently, a magnon induced transparency window emerges where the pump field is strongest (antinodes), and absorption occurs where the pump field is weakest (nodes). When considering the influence of magnon-phonon interaction, an additional transition path $\Ket{N_{a}, N_{m}+1, N_{b}}\leftrightarrow\Ket{N_{a}, N_{m}, N_{b}+1}$ introduces constructive interference, altering the transmission characteristics of the system.

\section{Possible experimental realization}
This section describes a potential practical implementation of our proposal and cites evidence from experiments that validate our theoretical analysis. We have considered a microwave cavity with a YIG sphere inside of it. The microwave cavity is generated by the fabrication of high-conductivity copper. The 250-micrometer-diameter YIG sphere operates as both the phonon and magnon resonator, with a phonon frequency of $\omega_{b} \approx 2\pi\times10$ MHz and a coupling strength of $g_{mb}\leq 2\pi\times3$ MHz \cite{zhang2016cavity}.

While the magnon is stimulated at gigahertz frequencies, the phonon mode is parametrically activated, that is, at the beating frequencies (megahertz) of the magnon modes. Because of its exceptional material as well as geometrical characteristics, the YIG sphere is also a great mechanical resonator. The phonon along magnon modes is connected to the changing magnetization brought about by the magnon excitation, which deforms the YIG sphere's spherical shape as well as vice versa.

The choice of cavity structure in cavity magnomechanics examines is important since it affects the capabilities of the experimental equipment. 3D microwave copper cavities are typically made from bulk metal (such as copper) and are designed in three dimensions, resembling a box-like structure. They often have precise dimensions to support resonant modes at microwave frequencies \cite{zhang2016cavity}. 3D cavities typically have well-defined input and output ports. These ports are crucial for coupling microwave signals into and out of the cavity. They can provide good isolation from external electromagnetic interference, which is important for sensitive measurements in magnomechanics.

However, planar cavities are fabricated on a flat surface, often using thin-film deposition techniques or microfabrication processes. They can be patterned to create resonant structures similar to those in 3D cavities. Planar cavities can also have defined input and output ports, though their positioning and geometry are typically different from 3D cavities. The ports may be integrated differently into the planar structure. Planar structures allow for miniaturization and integration with other components on a chip, which can simplify experimental setups and enable compact devices. Fabrication techniques offer precise control over the cavity dimensions and properties, potentially enhancing device performance.

Achieving efficient coupling of microwave signals into and out of planar cavities can be challenging, depending on the design and fabrication quality. Planar structures may be more susceptible to external electromagnetic interference compared to 3D cavities. The choice of cavity structure can influence the frequency range over which the magnomechanical interactions are studied. Different structures may support different modes and resonances. Planar structures offer advantages in terms of integration with other components such as magnetic materials or sensors, potentially enhancing device functionality.

\section{Conclusions}
\label{Sec:conclusions}
In this study, we introduced the MMIG in a magnomechanical cavity, featuring the interaction of a weak probe field with a standing wave control field applied to the cavity. The direct application of a magnetic field on a magnon induces magnon vibrations, subsequently leading to the generation of photon modes. Our analysis focused on the output spectrum of the probe light beam, considering the effects of cavity magnon interaction and magnon-phonon interaction. We explored the modulation in the transmission profile of the probe light beam under varying cavity magnon interaction strengths. The investigation revealed the influence of cavity magnon interaction, standing wave field strength, and interaction length on the transfer of probe energy into higher diffraction orders. Notably, our results could have potential applications in information storage and retrieval, particularly in implementing quantum memories with different orders of diffraction grating. These findings open avenues for further research and practical implementations in quantum information processing.

\appendix*
\section{APPENDIX: DERIVATION OF MAXWELL-BLOCH EQUATION}
We now briefly explain the propagation function of the probe light. Following the Maxwell-Bloch equation, the corresponding wave equation of incident probe field can be written as
\begin{equation}\label{A1}
    -\frac{\partial^{2}\mathcal{E}_{\text{p}}(z,t)}{\partial z^{2}}+\frac{1}{c^{2}}\frac{\partial^{2}\mathcal{E}_{\text{p}}(z,t)}{\partial t^{2}}=-\mu_{0}\frac{\partial^{2}P(z,t)}{\partial t^{2}}
\end{equation}
it can then be factorized 
\begin{equation}\label{A2}
    (\frac{\partial}{\partial z}+\frac{1}{c}\frac{\partial}{\partial t})(-\frac{\partial}{\partial z}+\frac{1}{c}\frac{\partial}{\partial t})\mathcal{E}_{\text{p}}(z,t)=-\mu_{0}\frac{\partial^{2}P(z,t)}{\partial t^{2}}
\end{equation}
where the corresponding polarization vector is $P(z,t)=\frac{1}{2} p(z,t)e^{-i(vt-kz)}+c.c$, with $\varepsilon_{\text{p}}(z,t)$ and $p(z,t)$ being carefully varied depending on time as well as positioning. The electric field vector for probing field is $\mathcal{E}_{\text{p}}(z,t)=\frac{1}{2} \varepsilon_{\text{p}}(z,t)e^{-i(vt-kz)}+c.c$.
When the change of $\varepsilon_{\text{p}}(z,t)$ and $p(z,t)$ in an optical frequency period are not obvious, we can apply slowly varying approximation:
\begin{equation}\label{A3}
    \frac{\partial  \varepsilon_{\text{p}}}{\partial t}<<v  \varepsilon_{\text{p}},\frac{\partial  \varepsilon_{\text{p}}}{\partial z}<<k  \varepsilon_{\text{p}},
    \frac{\partial  p}{\partial t}<<v  p,\frac{\partial  p}{\partial z}<<k  p.
\end{equation}
Under these approximations the propagation function of probe field reduce to
\begin{equation}\label{A4}
    ik\frac{\partial \mathcal{E}_{\text{p}}}{\partial z}+\frac{ik}{c}\frac{\partial \mathcal{E}_{\text{p}}}{\partial t}=-\frac{iv\mu_{0}}{2}[\frac{\partial P}{\partial t}-ivP]
\end{equation}
and under steady state condition the Eq. (A4) reduce to
\begin{equation}\label{A5}
    \frac{\partial \mathcal{E}_{\text{p}}}{\partial z}=-\frac{i \pi}{\lambda \varepsilon_0}(P)  
\end{equation}
The polarization vector $P$ is related to the probe field $\mathcal{E}_{\text{p}}$ via $\mathcal{E}_{\text{T}}$ as the relation $P=\varepsilon_{0} \mathcal{E}_{\text{T}} \mathcal{E}_{\text{p}}$.
By noting $\eta(\text{x})=(\frac{2\pi}{\lambda})\text{Re}[\mathcal{E}_{\text{T}}]$ and $\zeta(\text{x})=(\frac{2\pi}{\lambda})\text{Im}[\mathcal{E}_{\text{T}}]$,
the Eq. (A5) can be written as
\begin{align}
    \frac{\text{d}\mathcal{E}_{\text{p}}}{\text{d}z}=&[-\eta(\text{x})+\text{i}\zeta(\text{x})]\mathcal{E}_{\text{p}}\label{A6}
\end{align}
\label{appendix}
\section*{Acknowledgments}
Pei Zhang acknowledges the support of the National Nature Science Foundation of China (Grant No. 12174301), the Natural Science Basic Research Program of Shaanxi (Grant No. 2023-JC-JQ-01), and the Open Fund of State Key Laboratory of Acoustics (Grant No. SKLA202312).\\
\textbf{Data Availability Statement} This manuscript has associated data in a data repository.  All data included in this paper are available upon request by contacting with the corresponding author.
\bibliographystyle{apsrev4-2}
\bibliography{biblio}

\end{document}